\documentclass{iaus}
\usepackage{graphicx}

\title[JD 11.~~Metal-Poor Star Formation] 
{Dust and Gas as Seeds for Metal-Poor Star Formation}

\author[Deidre A. Hunter]   
{Deidre A. Hunter}

\affiliation{Lowell Observatory, 1400 West Mars Hill Road, Flagstaff, Arizona, USA
\\ email: {\tt dah@lowell.edu} 
}

\pubyear{2008}
\volume{255}  
\pagerange{1--11}
\jname{Low-Metallicity Star Formation: From the First Stars to Dwarf Galaxies}
\editors{L.K. Hunt, S. Madden \& R. Schneider, eds.}
\begin{document}

\maketitle

\begin{abstract}
I address the issue of dust and gas as seeds for metal-poor
star formation by reviewing what we know about star formation
in nearby dwarf galaxies and its relationship to the gas and
dust. I (try to) speculate on the extent to which processes
in nearby galaxies mimic star formation in the early universe.
\end{abstract}

\firstsection 
\section{Star formation and gas}

\begin{figure}[t]
\begin{center}
\includegraphics[width=2.0in]{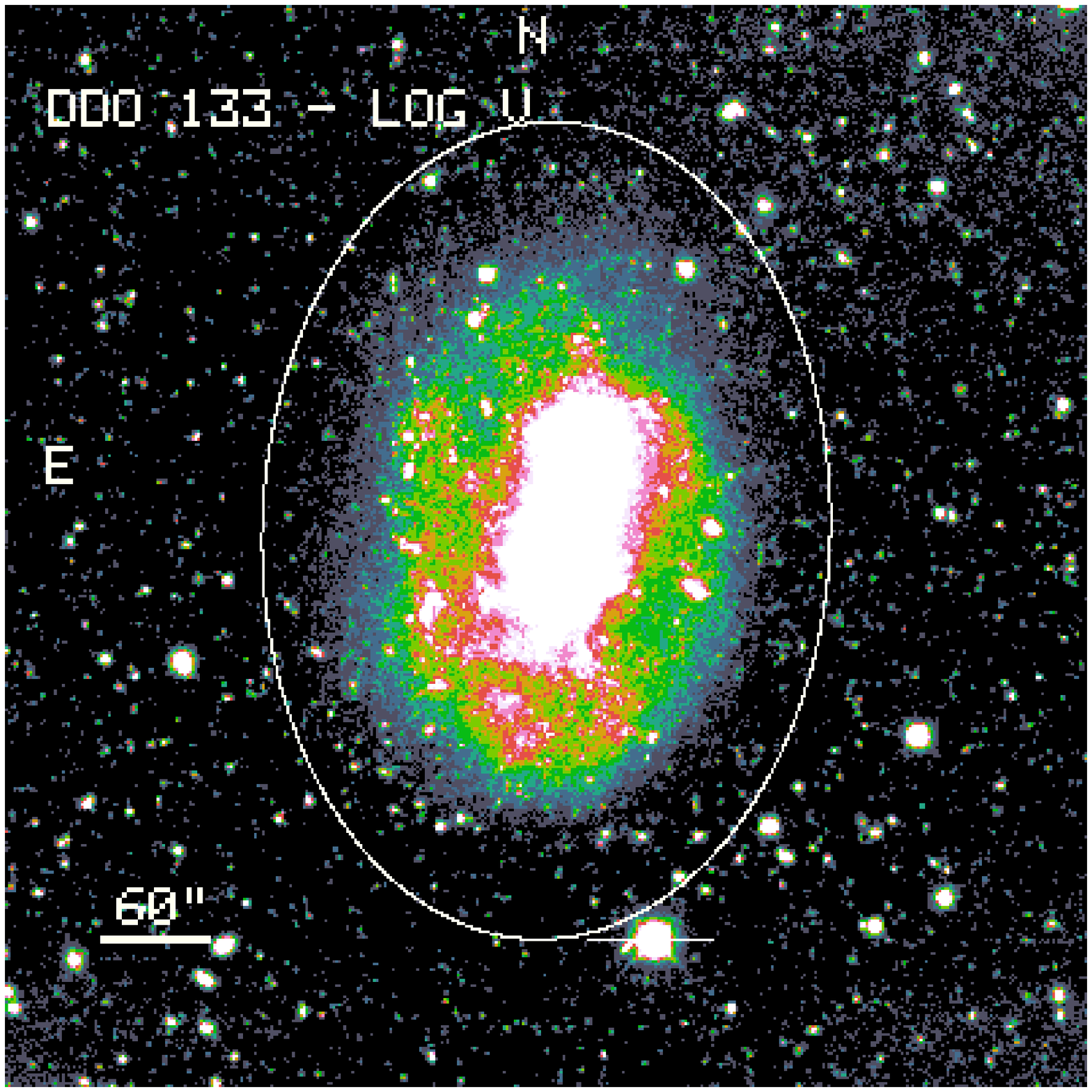}
\vspace*{-0.15 cm}
\includegraphics[width=1.75in]{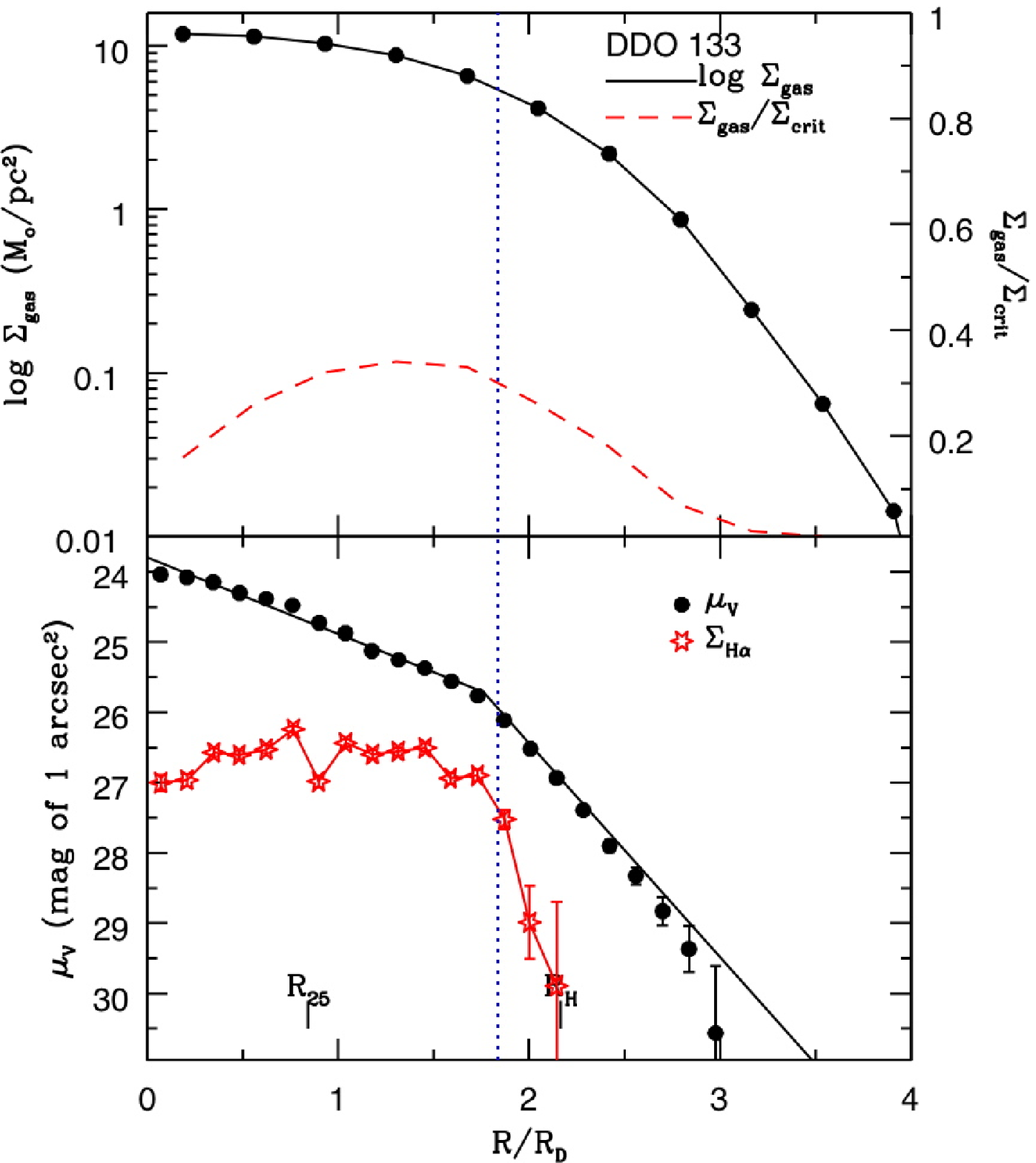}
\caption{{\it Left}: $V$-band image of the dIm DDO 133. 
{\it Right, top}: Gas surface density of DDO 133 and $\Sigma_{gas}/\Sigma_{crit}$. The
gas density is much lower than the Toomre critical density.
{\it Right, bottom}:  $V$-band and H$\alpha$ surface brightness profiles.
The $V$ profile is steeper in the outer disk.
} 
\label{fig-d133sb}
\end{center}
\end{figure}

Star formation in dwarf galaxies (dIm) is occuring at low, and even very low,
average atomic gas densities. The gas densities are sufficiently low that 
star formation is at best marginally unstable to large scale gravitational 
instabilities,
and that instability criterion does not predict where stars form
(models: \cite[Safronov 1960]{Safronov60}, 
\cite[Toomre 1964]{Toomre64}, \cite[Quirk 1974]{Quirk74}; 
dwarf observations:
\cite[Hunter \& Plummer 1996]{HunterPlummer96}, 
\cite[Meurer et al.\ 1996]{Meurer_etal96},
\cite[van Zee et al.\ 1997]{vanZee_etal97},
\cite[Hunter et al.\ 1998]{Hunter_etal98}, 
\cite[Rafikov 2001]{Rafikov01},
\cite[Leroy et al.\ 2008]{Leroy_etal08}). 
(See, for example, Figure \ref{fig-d133sb}, right top.)
This suggests that cloud formation in dwarfs is more difficult and inefficient than in the inner
parts of spirals 
(\cite[Dong et al.\ 2003]{Dong_etal03}, 
\cite[Li et al.\ 2005a]{Li_etal05a})
and that other, more local, processes dominate
(\cite[Elmegreen \& Hunter 2006]{ElmegreenHunter06}).
Furthermore, we expect the difficulty in forming stars to increase as the galaxy mass 
decreases 
(\cite[Li et al.\ 2005b]{Li_etal05b}).
So, just what do we know about star formation in dwarfs and its relationship to 
the (atomic) gas?

{\underline{\it Star formation is a local process}}. 

\begin{figure}[t]
\begin{center}
\includegraphics[width=2.0in]{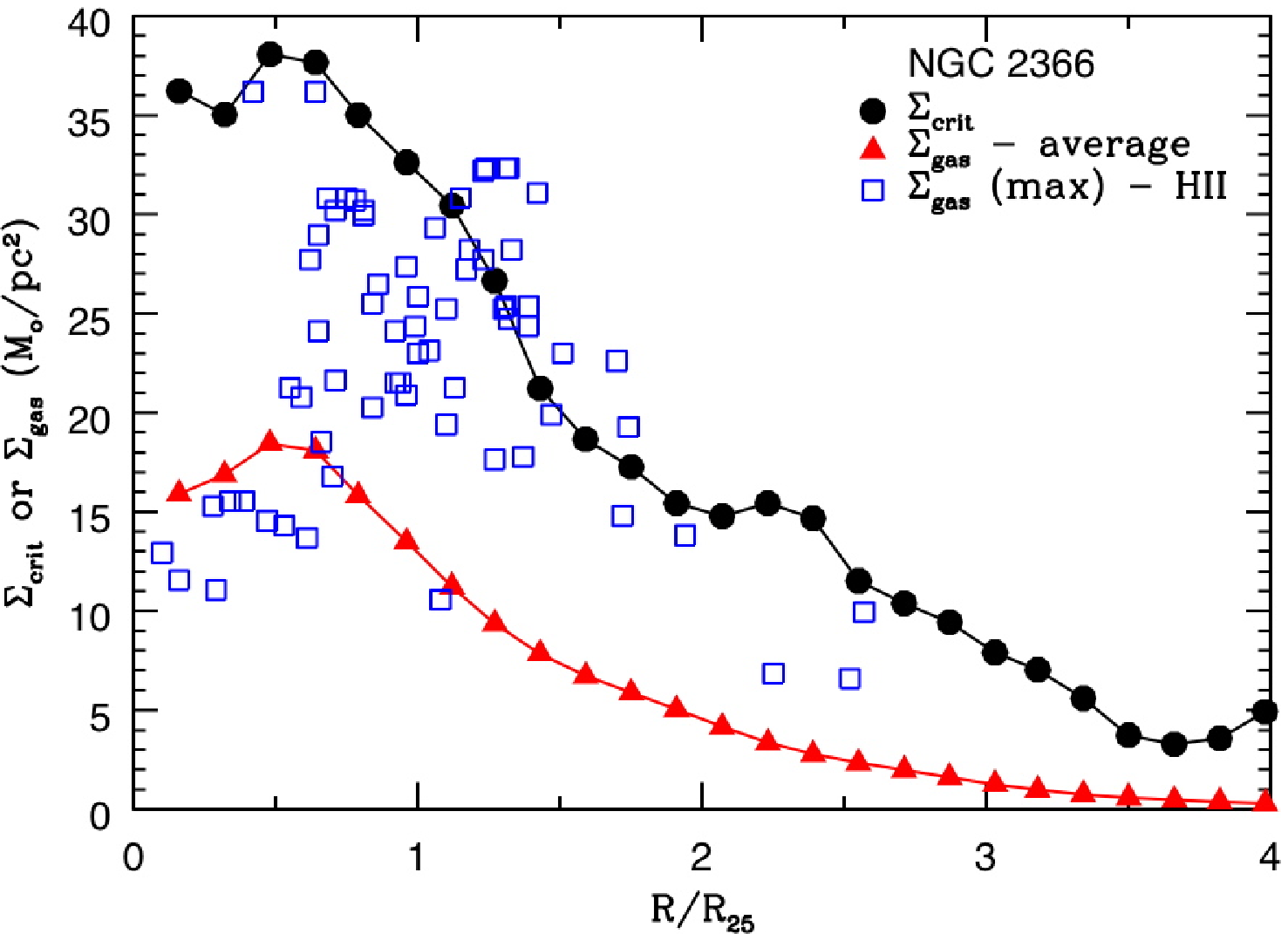}
\vspace*{-0.15 cm}
\includegraphics[width=2.5in]{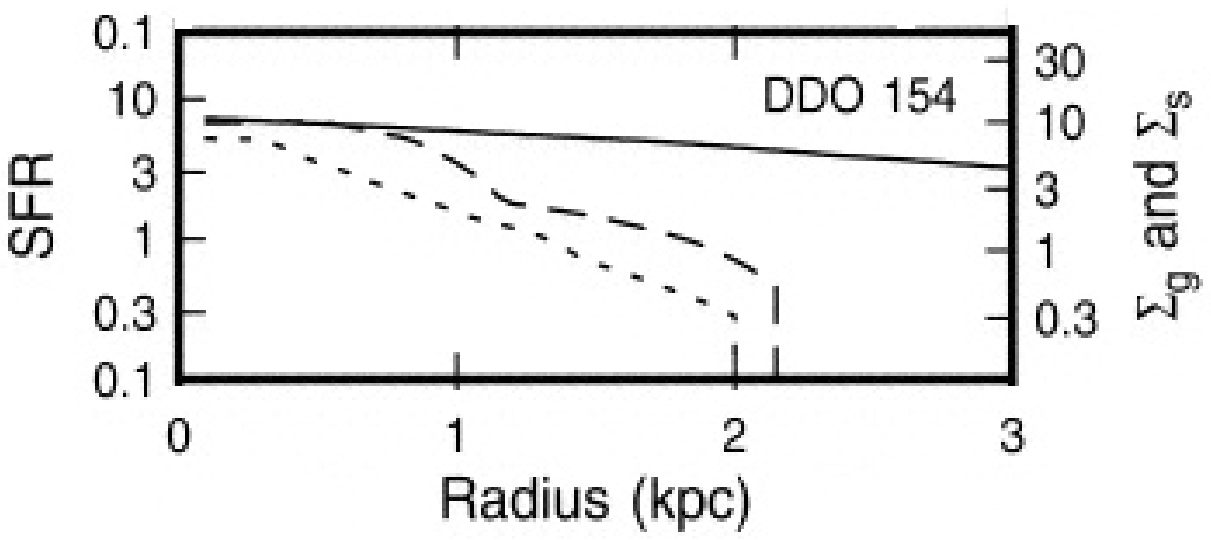}
\caption{{\it Left}: HI of dIm NGC 2366. The gas peaks in star-forming regions
are close in density to the Toomre critical density even though the average
surface density is much lower. Adapted from \cite[Hunter et al.\ (2001a)]{Hunter_etal01a}.
{\it Right}: Example of star formation rate profile ({\it long dashed line}) following the 
average surface density of older stars ({\it short dashed line}) better than it follows the average
gas surface density ({\it solid line}). Adapted from 
\cite[Hunter et al.\ (1998)]{Hunter_etal98}.
} 
\label{fig-n2366}
\end{center}
\end{figure}

Since large-scale spontaneous gravitational instabilities are not dominant,
other---local---processes must be important in tiny galaxies.
That this is the case is seen observationally: 
Star formation is occurring in HI clouds or complexes 
even where the {\it average} gas density is ``too low''
(\cite[Hunter \& Gallagher 1985, 1986]{HunterGallagher85}; 
\cite[Phillipps et al.\ 1990]{Phillipps_etal90}; 
\cite[van der Hulst et al.\ 1993]{vanderHulst_etal03};
\cite[Taylor et al.\ 1994]{Taylor_etal94}; 
\cite[van Zee et al.\ 1997]{vanZee_etal97};
\cite[Meurer et al.\ 1998]{Meurer_etal98}; 
\cite[de Blok \& Walter 2006]{deBlokWalter06}).
For example, in NGC 2366 the HI peaks associated with star-forming regions
are close to the Toomre critical density even though the average gas density
in the surroundings is well below it 
(see Figure \ref{fig-n2366}; \cite[Hunter et al.\ 2001a]{Hunter_etal01a}).
Star formation in outer disks, where the gas is highly sub-critical, must certainly
be a local process 
(\cite[Roye \& Hunter 2000]{RoyeHunter00}, 
\cite[Komiyama et al.\ 2003]{Komiyama_etal03}, 
\cite[Parodi \& Binggeli 2003]{ParodiBinggeli03}).

{\underline{\it Not all HI is equal}}. 

However, not all of the HI is equal when it comes to star formation.
\cite[Braun (1997)]{Braun97} 
found that 60--90\% of HI is in cool filaments in spirals and dwarfs (but see Usero et al., in prep),
and it is the cool HI that we expect to be associated with new star formation.
In fact, 
\cite[Young \& Lo (1996)]{YoungLo96} 
found that
20\% of the HI in Leo A is in a component with a low velocity dispersion, and 
similarly for a small sample of other dwarfs 
(\cite[Young et al.\ 2003]{Young_etal03}, 
\cite[Begum et al.\ 2006b]{Begum_etal06b}). 
Even little Leo T, the recently discovered dIm in the Local Group with
an M$_V$ of only $-7$ 
(\cite[Irwin et al.\ 2007]{Irwin_etal07}),
has a cool HI component ($\sigma\sim 2$ km/s;
\cite[Ryan-Weber et al.\ 2008]{Ryan_etal08}).
In NGC 6822, 
\cite[de Blok \& Walter (2006)]{deBlokWalter06} 
concluded that the cool gas is more important
in determining the star formation {\it locally} than the total HI, although the relationship
between cool HI and star formation is not deterministically simple 
(\cite[Young et al.\ 2003]{Young_etal03},
\cite[Begum et al.\ 2006b]{Begum_etal06b}).
Thus, the immediate reservoir of gas for cloud formation may not be as extensive
as the integrated HI mass would indicate.

{\underline{\it Star formation follows the older stars}}. 

The azimuthally-averaged star formation rate follows the stellar surface density
of the older stars
with radius better than it follows the gas surface density in most dwarfs
(see Figure \ref{fig-n2366}). 
So, star formation 
appears to be tied to existing stars
(\cite[Hunter \& Gallagher 1985]{HunterGallagher85}, 
\cite[Brosch et al.\ 1998]{Brosch_etal98}, 
\cite[Hunter et al.\ 1998]{Hunter_etal98},
\cite[Stewart et al.\ 2000]{Stewart_etal00},
\cite[Parodi \& Binggeli 2003]{ParodiBinggeli03}, 
\cite[Hunter \& Elmegreen 2004]{HunterElmegreen04},
\cite[Leroy et al.\ 2008]{Leroy_etal08}), 
with some time averaging
($10^8$ yrs, \cite[Andersen \& Burkert 2000]{AndersenBurkert00}).
In many galaxies, this is true even in the far outer disk.

{\underline{\it Stars blow holes, and porosity has consequences}}. 

One way for star formation to be tied to the presence of older stars is through
star-induced star formation 
(\cite[\"Opik 1953]{Opik53}; 
\cite[Gerola et al.\ 1980]{Gerola_etal80}; 
\cite[Comins 1983]{Comins83}).
Star formation changes the interstellar medium (ISM; 
\cite[Martin 1997]{Martin97}, 
\cite[M\"uhle et al.\ 2005]{Muhle_etal05}).
Massive stars have strong winds and then die as supernovae, so where
concentrations of massive stars form, large holes can be blown in the ISM
(\cite[Puche et al.\ 1992]{Puche_etal92},
\cite[Martin 1998]{Martin98},
\cite[Walter \& Brinks 1999]{WalterBrinks99},
\cite[Parodi \& Binggeli 2003]{ParodiBinggeli03}, 
\cite[Cannon et al.\ 2005]{Cannon_etal05}, 
\cite[de Blok \& Walter 2006]{deBlokWalter06}).
These holes are surrounded by higher density shells and in these shells
star-forming clouds can develop 
(\cite[Mori et al.\ 1997]{Mori_etal97}, \cite[van Dyk et al.\ 1998]{vanDyk_etal98}).
A  beautiful example is Constellation III in the LMC 
(\cite[Dopita et al.\ 1985]{Dopita_etal85}). 
There, stars that formed in the interior
12--16 Myr ago blew a 1.4-kpc size hole in the gas   
(\cite[Dolphin \& Hunter 1998]{DolphinHunter98}). 
More recently ($\sim$6 Myrs) stars have formed in
the higher density shell surrounding the hole.

\begin{figure}[t]
\begin{center}
\includegraphics[width=1.5in]{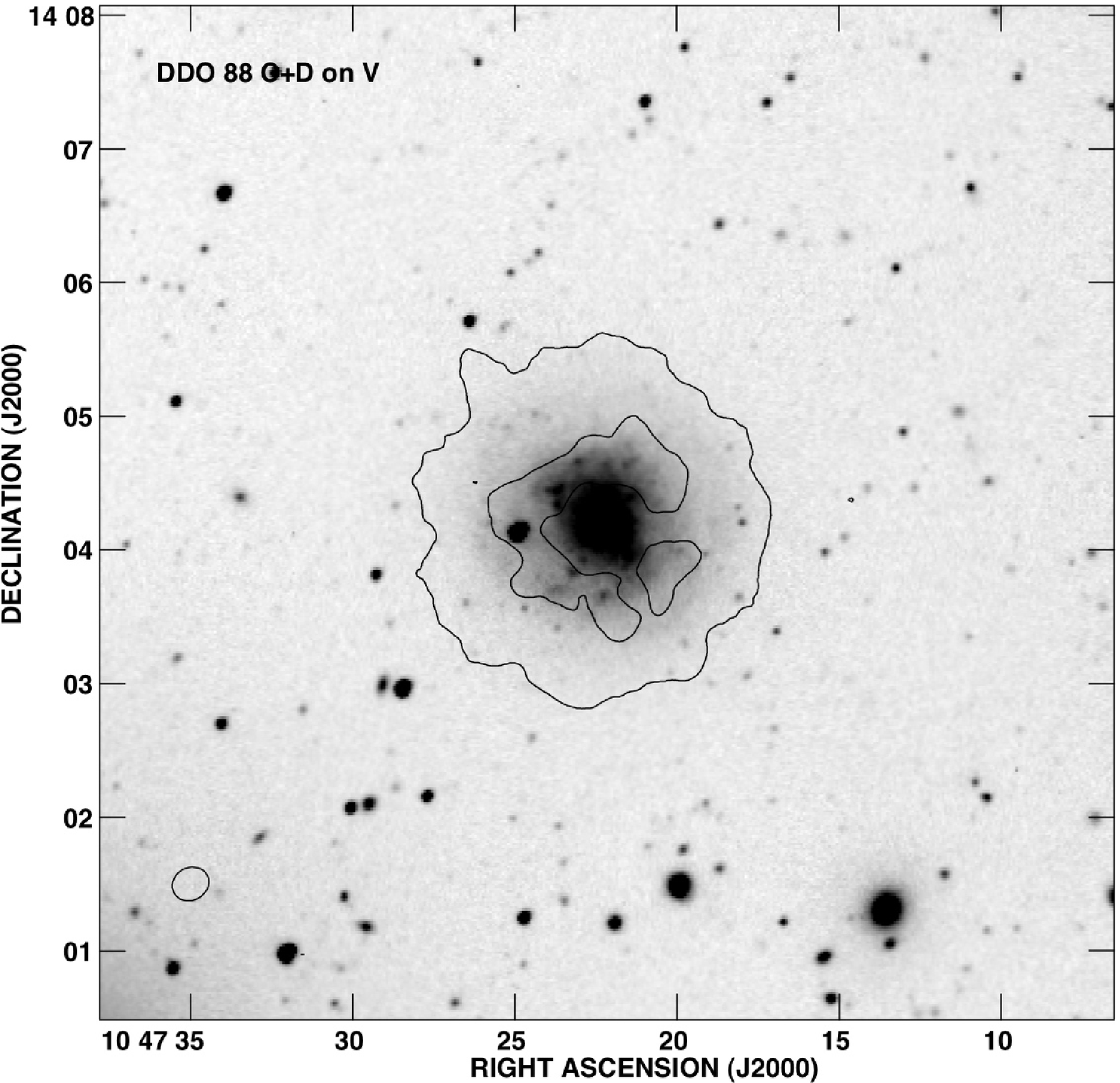}
\vspace*{-0.15 cm}
\includegraphics[width=2.5in]{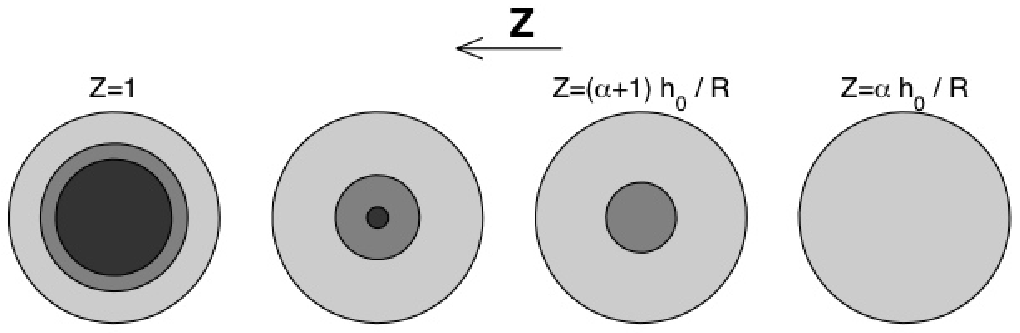}
\caption{{\it Left}: $V$ image of the Sm DDO 88 with contours of integrated HI emission
(600$\times$520 pc beam;  \cite[Simpson et al.\ 2005]{Simpson_etal05}). 
Stars have blown a 3-kpc diameter gas ring.
{\it Right}: Sketch of molecular clumps as a function of metallicity from
\cite[Bolatto et al.\ (1999)]{Bolatto_etal99}. As the metallicity goes down, the molecular
core traced by CO shrinks and the PDR grows.
}
\label{fig-d88}
\end{center}
\end{figure}

The action of the massive stars can
create a swiss-cheese morphology in the ISM, and some dwarfs have lots of
holes in their gas.
In fact the filling factor in holes can be very high, exceeding that in M31 by factors
of 3--800 
(\cite[Oey et al.\ 2001]{Oey_etal01};
Bagetakos et al., in prep).
The rearrangement of the ISM can also be severe in tiny galaxies, and without shear,
structures can last a long time.
In DDO 88, for example, the HI looks like a giant doughnut and the ring, located
at $1/2 R_H$, contains 30\% of the total HI mass 
(see Figure \ref{fig-d88}; \cite[Simpson et al.\ 2005]{Simpson_etal05}).
The massive stars may even blow the gas out of the galaxy, and completely away
if the total mass is small enough, about $10^7$ M$_{\mathord\odot}$
(\cite[Martin 1998]{Martin98}, 
\cite[Mac Low \& Ferrara 1999]{MacLowFerrara99}, 
\cite[Ferrara \& Tolstoy 2000]{FerraraTolstoy00},
but see 
\cite[Ott et al.\ 2005]{Ott_etal05}).

Not all holes have to be made by massive stars. 
There are holes in outer disks and low density HI regions
without obvious young stars 
(LMC: \cite[Kim et al.\ 1999]{Kim_etal99}; 
Holmberg II: \cite[Rhode et al.\ 1999]{Rhode_etal99}, 
\cite[Stewart et al.\ 2000]{Stewart_etal00}, 
\cite[Bureau \& Carignan 2002]{BureauCarignan02}; 
DDO 154: \cite[Hoffman et al.\ 2001]{Hoffman_etal01}), 
and models show that these could
be the result of turbulence and other processes such as thermal and gravitational instabilities
(\cite[Wada et al.\ 2000]{Wada_etal00}, 
\cite[Piontek \& Ostriker 2004]{PiontekOstriker04},
\cite[Dib \& Burkert 2005]{DibBurkert05}, 
but see \cite[S\'anchez-Salcedo 2001]{Sanchez01}).

But what are the consequences of this porosity to star formation? 
Although secondary star formation can form in the shells,
porosity in general may also regulate the star formation process by
making it harder to form more stars---like a speed bump to star formation
(\cite[Silk 1997]{Silk97}, 
\cite[Scalo \& Chappell 1999]{ScaloChappell99},
\cite[Martin 1999]{Martin99}, 
\cite[Parodi \& Binggeli 2003]{ParodiBinggeli03}). 
This is partly because the holes allow UV photons to heat the gas and reduce
the cold gas supply 
(\cite[Silk 1997]{Silk97}). 
In addition, one can imagine that the porosity
would make it hard to form multiple generations beyond the second one.
For example, in the LMC, the HI is so filamentary that it is hard to see how the 
second generation in the shell of Constellation III is going to have much of a chance
to form a third generation 
(see the HI map of \cite[Kim et al.\ 1999]{Kim_etal99}).
For the galaxy as a whole, which process---secondary star formation
or regulating further star formation---wins or the roles they both play aren't clear.

One consequence of this on-going rearrangement of the ISM
is that star formation must move around the galaxy on kpc scales.
And this is what is observed 
(\cite[Payne-Gaposhkin 1974]{Payne74}; 
\cite[Hodge 1969, 1980]{Hodge69}; 
\cite[Hunter et al.\ 1982]{Hunter_etal82};
\cite[Hunter \& Gallagher 1985]{HunterGallagher85},
\cite[Hunter \& Gallagher 1986]{HunterGallagher86},
\cite[Schombert et al.\ 2001]{Schombert_etal01}).
These statistical sorts of fluctuations across a galaxy
can be severe in dwarfs just because they are small.
For example, you could end up with a galaxy that is blue in one half, where star formation
took place recently, and red in the other half, where older stars dominate
(\cite[Hunter \& Elmegreen 2006]{HunterElmegreen06}).

{\underline{\it There is evidence for turbulence compression}}. 

There is evidence for another local process taking place in dwarfs---turbulence
compression.
First, the ISM is structured into clouds of all sizes (that is, a fractal) whose distributions
resemble those of compressible turbulence. This is seen in the LMC
(\cite[Elmegreen et al.\ 2001]{Elmegreen_etal01}), the SMC 
(\cite[Stanimirovic et al.\ 1999]{Stanimirovic_etal99}), and other nearby dwarfs 
(\cite[Westpfahl et al.\ 1999]{Westpfahl_etal99}, 
\cite[Begum et al.\ 2006a]{Begum_etal06a}).
Second, other distributions are also consistent with sampling a fractal turbulent gas,
including the HII region luminosity function 
(\cite[Kingsburgh \& McCall 1998]{KingsburghMcCall98}, 
\cite[Youngblood \& Hunter 1999]{YoungbloodHunter99}),
H$\alpha$ probability distribution function 
(\cite[Hunter \& Elmegreen 2004]{HunterElmegreen04}),
$V$-band Fourier Transform power spectra 
(\cite[Willett et al.\ 2005]{Willett_etal05}), and
star cluster mass functions 
(e.g., \cite[Hunter et al.\ 2003]{Hunter_etal03}).
In Holmberg II, for example, 
\cite[Dib \& Burkert (2005)]{DibBurkert05} see correlated structure in the
ISM for scales less than 6 kpc and argue for a turbulence driver that acts on scales of order 6 kpc.

Turbulence can heat the gas and make it harder to form stars, thereby regulating star formation
(\cite[Struck \& Smith 1999]{StruckSmith99}).
 But it can also bring gas together
and create density enhancements that become self-gravitating if the gas is dense enough for the gravitational potential energy to exceed the energy in turbulent motions 
(\cite[Krumholz \& McKee 2005]{KrumholzMcKee05}).

{\underline{\it Stars form in outer disks}}. 

Stars have formed in outer disks in the past, and 
FUV surface brightness profiles indicate that, in many dwarfs, young stars also exist far into
the outer disk 
(Hunter et al., in prep). 
Ultra-deep
imaging shows that the stellar disks continue to surface brightnesses of even
30 mag/arcsec$^2$ in $V$ with no end in sight 
(see Figure \ref{fig-d133sb}).
Thus, stars have formed in extreme low-density conditions 
in outer dwarf disks, and this region poses a particularly difficult test of our
understanding of the cloud/star formation process.

In outer disks, we often find complex stellar surface brightness profiles (see Figure \ref{fig-d133sb}).
About 24\% of a survey of 94 dIm show a stellar surface brightness profile that becomes
suddenly steeper at about 2 disk scale lengths 
(\cite[Hunter \& Elmegreen 2006]{HunterElmegreen06}).
This kind of break is also seen in spirals 
(see, for example, \cite[Pohlen et al.\ 2002]{Pohlen_etal02},
\cite[Kregel \& van der Kruit 2004]{KregelvanderKruit04}) 
and in disks at redshifts of $0.6<z<1.0$ 
(\cite[P\'erez 2004]{Perez04}).
People have historically called this a ``truncation'' because they thought they were seeing
the end of the disk, but I think it is better referred to as a break because it represents
a change rather than an ending. The difference between spirals and dwarfs is that the
break occurs much closer into the center of the galaxy in dwarfs.

\begin{figure}[t]
\begin{center}
\includegraphics[width=1.75in]{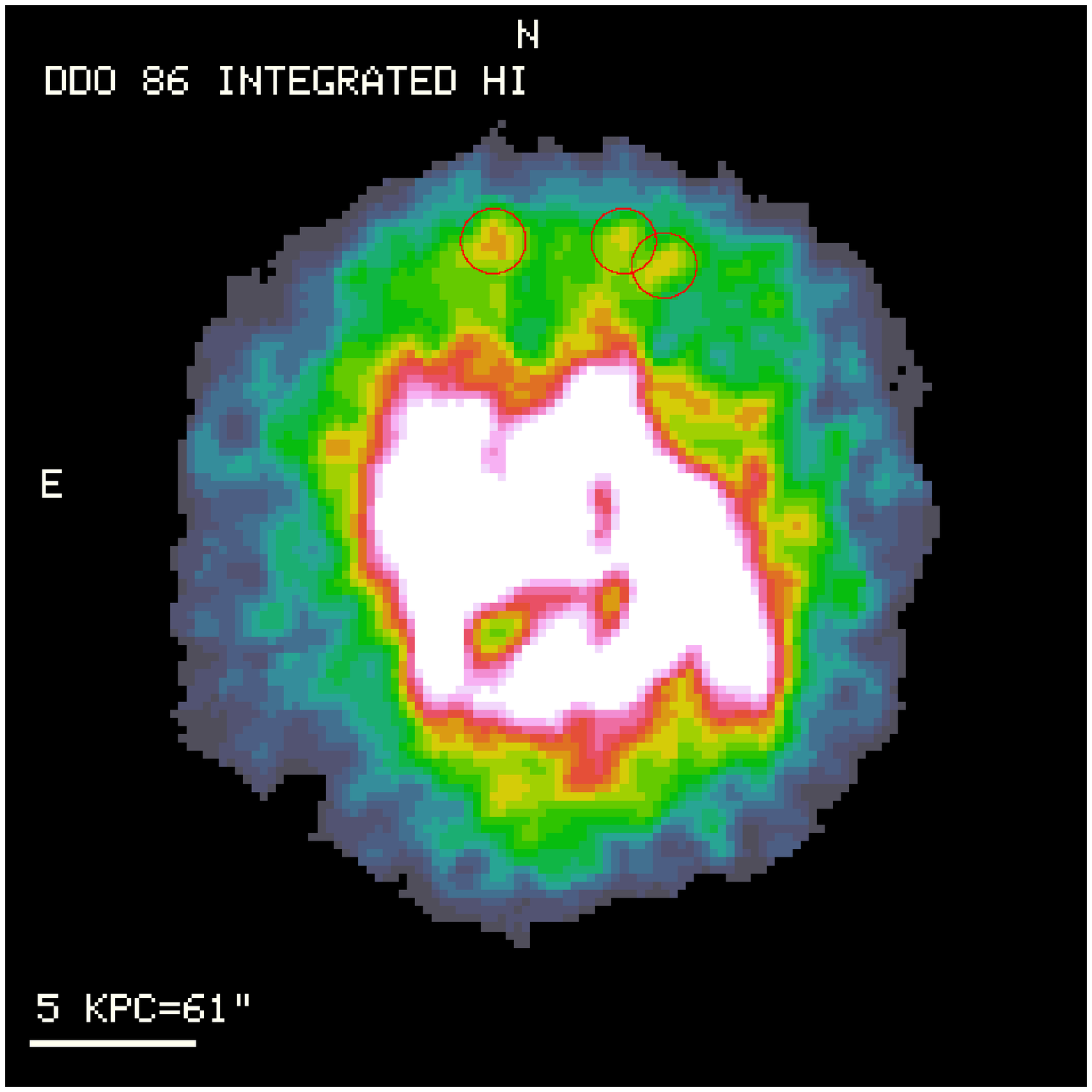}
\vspace*{-0.15 cm}
\includegraphics[width=2.8in]{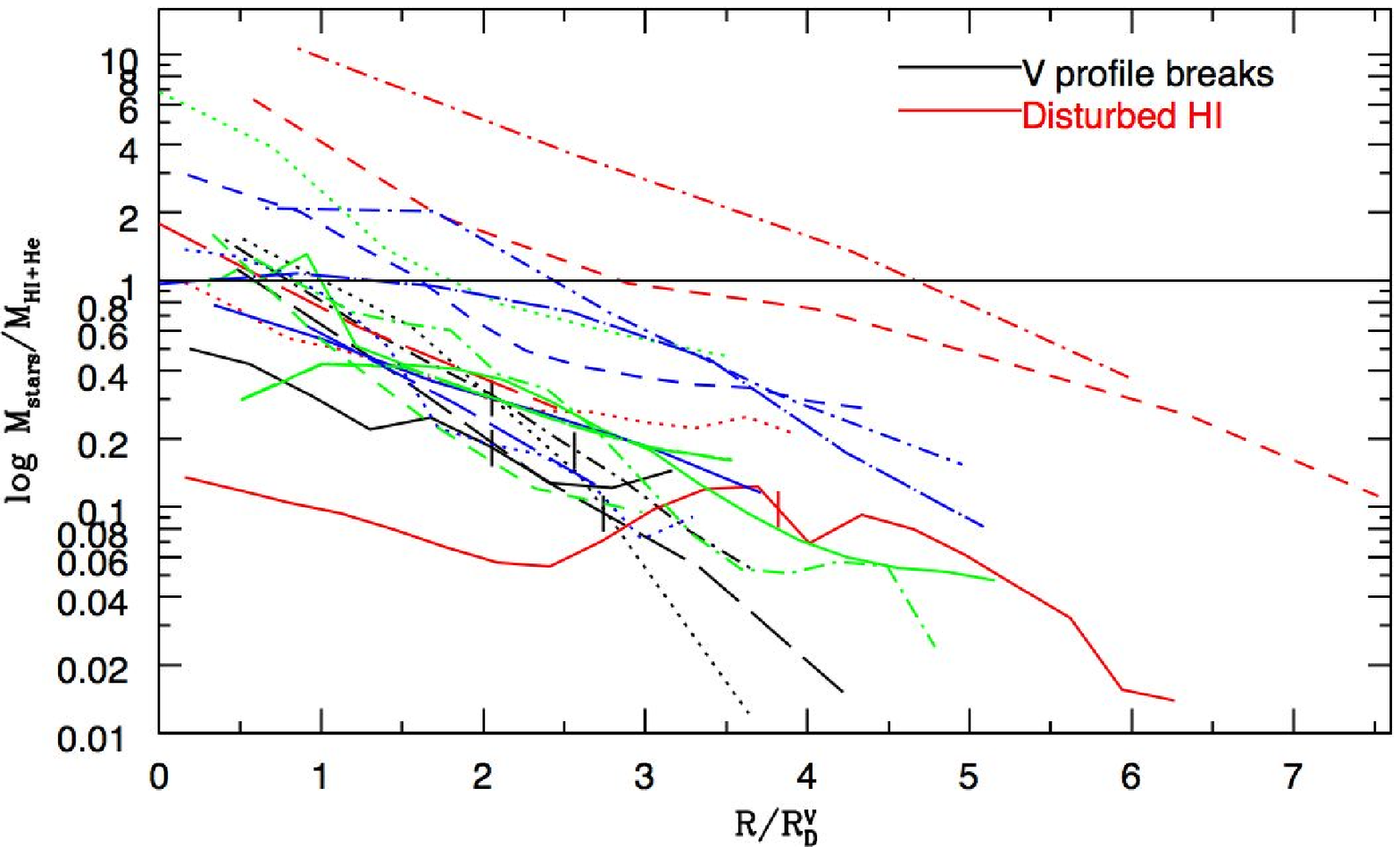}
\caption{{\it Left}: Integrated HI map of dIm DDO 86 (1.4$\times$1.3 kpc beam).
Circled HI peaks at $5R_D$ have maxima of $5\times10^{20}$/cm$^2$, within expected thresholds
for star formation (Schaye 2004), even though the average
$\Sigma_{gas}$ there is $3\pm1\times10^{20}$/cm$^2$.
{\it Right}: $M_{gas}/M_{stars}$ for dwarfs. 
Dwarfs become increasingly gas-rich with radius, implying a decreasing 
cloud formation efficiency.
}
\label{fig-outerpeaks}
\end{center}
\end{figure}

What happens to the star formation process at the break?
Different theoretical approaches predict such breaks:
1) \cite[Andersen \& Burkert (2000)]{AndersenBurkert00} 
suggest that self-regulated evolution within a confining dark halo
will lead naturally to exponential density profiles that are somewhat flatter in the central regions.
2) \cite[Schaye (2004)]{Schaye04} 
argues that star formation occurs where there is cold gas, triggering 
gravitational instabilities. The threshold gas column density for this transition is around
3--10$\times10^{20}$/cm$^2$.
Tiny Leo T has a peak column density of 
7$\times10^{20}$/cm$^2$ and it has formed stars as recently as 200 Myr ago
(\cite[Ryan-Weber et al.\ 2008]{Ryan_etal08}, \cite[Irwin et al.\ 2007]{Irwin_etal07}).
And star formation is found where locally the gas density is greater than 4-6$\times10^{20}$/cm$^2$
in a sample of low luminosity dwarfs 
(\cite[Hunter et al.\ 2001a]{Hunter_etal01a}, \cite[Begum et al.\ 2006b]{Begum_etal06b}).
The break, according to Schaye, occurs where the {\it average} gas density drops below this
threshold.
3) \cite[Li et al.\ (2005b)]{Li_etal05b} 
determined from hydrodynamic simulations
that there should be a sharp drop in the star formation rate at 2$\times$ the disk scale length
and that stars are more important than gas in destabilizing dwarf disks.
4) \cite[Elmegreen \& Hunter (2006)]{ElmegreenHunter06} 
suggest that inner disks are dominated by large-scale
gravitational instabilities while in outer disks local processes driven by a low level of 
bulk motions of the gas form clouds with densities high enough {\it locally} to form stars
(see Figure \ref{fig-outerpeaks}). 
The profile break is the transition from the dominance of large-scale processes
to the dominance of local processes.
This requires a low level of motion of the gas continuing into the outer disk, as is observed.
One mechanism for producing local density enhancements 
is the magnetorotational instability, where the
angular velocity decreases outward and magnetic fields are present. 
\cite[Piontek \& Ostriker (2005)]{PiontekOstriker05} predict that turbulent velocity
dispersions go up, reaching a quasi-steady plateau, as the average gas density goes down.
This results in departures from thermal equilibrium and local density variations that
might be particularly useful in outer disks.
5) \cite[Ro\v skar et al.\ (2008)]{Roskar_etal08} 
suggest that the break forms in a (spiral) disk within 1 Gyr
and that the break moves outward as the gas cools. The break in their model
corresponds to a rapid drop in the star formation rate
associated with a drop in the cool gas surface density relative to the Toomre
critical density.
6) \cite[Bournaud et al.\ (2007)]{Bournaud_etal07} suggest the outer exponential is built from tidal debris.
All but the last model agree that the break corresponds to a change in the star formation
rate due to changes in the physical conditions within the gas.

{\underline{\it There is a wide variety of gas surface density profiles}}. 

One thing all dwarfs have in common is that the ratio of the mass 
in stars to mass in gas changes steadily with radius. 
Dwarfs are usually gas dominated even in the centers, but become more so with radius.
This implies a steady decrease in the large-scale cloud formation efficiency 
with radius
(see Figure \ref{fig-outerpeaks}; \cite[Leroy et al.\ 2008]{Leroy_etal08}).

The wide variety of gas surface density fall-offs with radius is striking.
Some profiles are relatively flat (in terms of log of the surface density), 
some drop precipitously,
and others decrease steadily.
So, just what is the role of the gas in determining the nature of the stellar disk, especially
outer disks?
To explore the relationship of the HI profiles to properties of the stellar disk,
I fit Sersic functions to HI surface density profiles of 19 dwarf galaxies:
$\log \Sigma_{gas}(R) = (\log \Sigma_{gas})_0 - 0.434 (R/R_0)^{1/n}$.
Dwarfs are fit well with $n\leq1$; a higher $n$ value means the gas is more centrally
concentrated.

\begin{figure}[t]
\begin{center}
\includegraphics[width=2.25in]{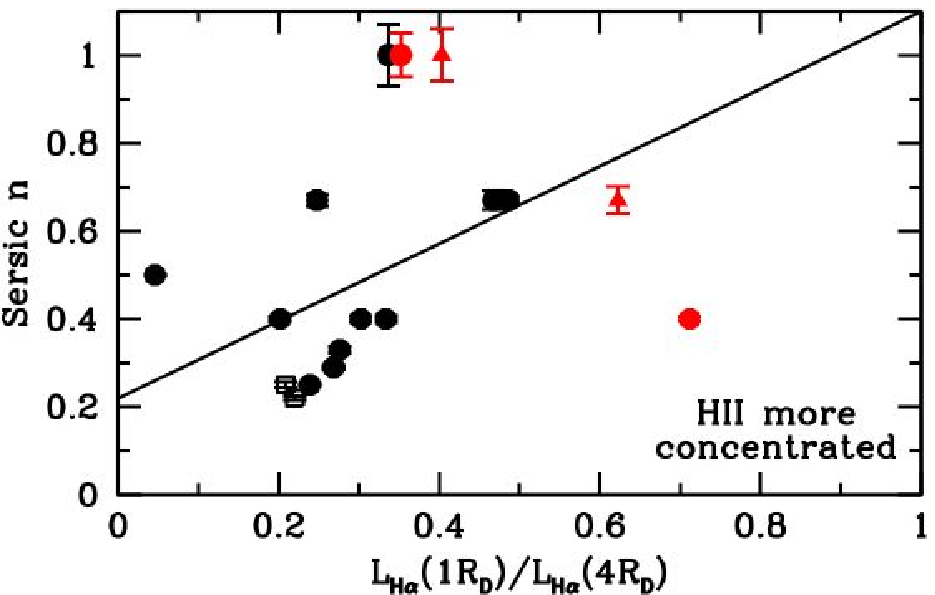}
\vspace*{-0.15 cm}
\includegraphics[width=2.15in]{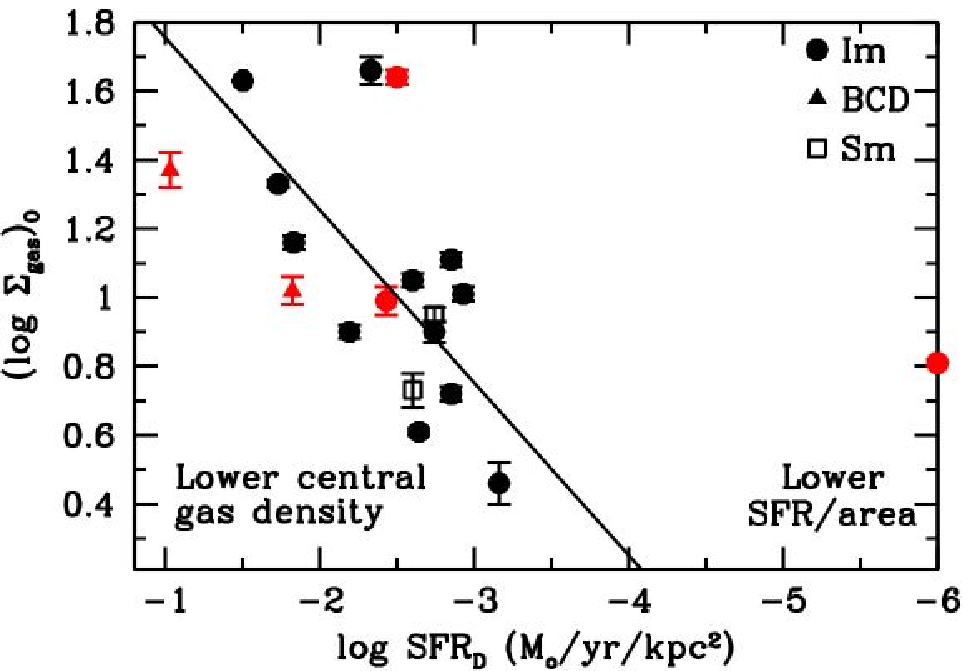}
\caption{Results of fitting gas surface density profiles with Sersic functions. 
{\it Left}: Sersic $n$ vs.
$L_H\alpha$(1$R_D$)/$L_H\alpha$(4$R_D$), a measure
of the central concentration of the star formation activity. 
A higher $n$ means  a higher degree of central concentration of HI.
The star formation activity is more centrally
concentrated in galaxies where the HI is also more centrally concentrated.
{\it Right}: The integrated star formation rate normalized to $\pi R_D^2$ SFR$_D$ correlates with 
the central gas density, not the shape of the $\Sigma_{gas}$ profile.
}
\label{fig-Sersic}
\end{center}
\end{figure}

There are a few trends between the way the gas falls off with radius and
the star-forming characteristics of the galaxies (see Figure \ref{fig-Sersic}).
First, the star formation activity is more centrally concentrated in galaxies where
the gas is more centrally concentrated.
Second, the central gas density is more important in determining the integrated star formation rate
than the details of the gas profile; a higher central gas density
correlates with a higher integrated star formation rate.
Third, galaxies with breaks in their stellar surface brightness profiles tend to have
gas profiles that are less centrally concentrated (lower $n$) and have lower central
gas densities, but there are similar galaxies without optical breaks as well.
So, all in all, the gas does seem to play a role in determining the stellar disk, but the effects
are not as dramatic as the variation in gas profiles.

{\underline{\it No dark galaxies}}. 

Almost all puddles of gas must form stars according to 
\cite[Taylor \& Webster (2005)]{TaylorWebster05}. 
In their models 
the gas in galaxies with more than $10^6$ M$_{\mathord\odot}$ in gas and stars
becomes unstable and forms stars until there is a stellar radiation field to warm the ISM and stabilize it. 
The key to instability is efficient H$_2$ cooling.
Similarly, \cite[Warren et al.\ (2007)]{Warren_etal07}
suggest that isolated
galaxies with shallow dark matter potentials will produce the minimum quantity of stars
needed to stabilize the gas disk.
Taylor and Webster suggest that
local dwarfs represent the minimum rates of self-regulated star formation. 
Leo T must define the extreme. It has only 3$\times10^5$ M$_{\mathord\odot}$
of HI and a peak column density of 7$\times10^{20}$/cm$^2$, but it has formed stars
in the past 200 Myrs
(\cite[Irwin et al.\ 2007]{Irwin_etal07}, \cite[Ryan-Weber et al.\ 2008]{Ryan_etal08}). 

{\underline{\it Starbursts happen}}. 

\cite[Lee et al.\ (2007)]{Lee_etal07} 
find evidence that the star formation histories of dwarfs with $M_B>-15$
are more erratic than those of brighter dwarfs. Specifically, the equivalent width of H$\alpha$ becomes
more scattered. They argue that at this brightness level there is some change
in the star formation process, or at least in the regulatory nature of the process. One possibility is
that feedback from massive stars has a larger negative impact, and the evolution becomes bursty
or at least ``gaspy'' 
(\cite[Marconi et al.\ 1995]{Marconi_etal95}, \cite[Schombert et al.\ 2001]{Schombert_etal01}).
This emphasizes the need to look for trends across the dwarf galaxy class.

\cite[Stinson et  al.\ (2007)]{Stinson_etal07} 
have simulated the collapse of dwarf galaxies. They find that supernovae
feedback can quench star formation as the gas collapses. Gas flows out in a hot halo, cools,  and
then forms stars again. The result is episodic bursts of star formation.
On the other hand, 
\cite[Li et al.\ (2005a)]{Li_etal05a} 
argue that direct feedback from starbursts 
is minor because most of the energy is deposited above the disk, not in it.

Blue Compact Dwarfs (BCDs) have, or are, undergoing enhanced star formation at some
level. Furthermore, star formation has migrated to the center within the last Gyr
(\cite[Noguchi 1988]{Noguchi88}, 
\cite[Hunter \& Elmegreen 2004]{HunterElmegreen04},
but see \cite[van Zee et al.\ 2001]{vanZee_etal01}).  
The HI is more centrally concentrated 
(\cite[Chamaraux 1977]{Chamaraux77}, 
\cite[Taylor et al.\ 1994]{Taylor_etal94}, 
\cite[Meurer et al.\ 1998]{Meurer_etal98}, 
\cite[van Zee et al.\ 2001]{vanZee_etal01}), 
and the distribution of HI has changed with time 
(\cite[Simpson \& Gottesman 2000]{SimpsonGottesman00}). 
Some have unusually large HII regions 
(\cite[Youngblood \& Hunter 1999]{YoungbloodHunter99}),
and many have peculiar HI velocity fields.
Sometimes there is some object nearby to blame for the starburst, but not always
(c.f., \cite[Taylor 1997]{Taylor97}; 
\cite[Stil \& Israel 1998]{StilIsrael98}; 
\cite[M\'endez \& Esteban 2000]{MendezEsteban00}; 
\cite[Campos-Aquilar \& Moles 1991]{CamposMoles91}; 
\cite[Telles \& Terlevich 1995]{TellesTerlevich95}; 
\cite[Telles \& Maddox 2000]{TellesMaddox00}; 
Simpson et al.\ 2008, in prep).
Could some BCDs be advanced dwarf--dwarf mergers or dwarf and even-smaller-dwarf mergers (so
there won't be a tidal tail, which we don't see)? 
BCDs may be showing us the frequency of dwarf--dwarf interactions.

\section{Star formation and dust}

The relationship between the gas and star formation is obvious since stars form out of
gas clouds. The connection between dust and star formation is less direct.
Dust affects the heating cycle in the ISM, and this affects the ability of the ISM to form cold, dense
clouds that can form stars. The presence of even a small amount of dust can make a
difference to the ability to form H$_2$ clouds
(\cite[Schaye 2004]{Schaye04}).
Thus, the low dust content {\it should} have consequences
to the star formation process in dwarfs.

{\underline{\it The dust and metal contents are low}}. 

Although there are embedded star-forming regions in dwarfs 
and even dust in outer disks
(e.g., \cite[Hinz et al.\ 2006]{Hinz_etal06}, 
\cite[Walter et al.\ 2007]{Walter_etal07}),
the dust-to-gas ratio is lower in dwarfs than in spirals by factors of 2--25
(\cite[Hunter et al.\ 1989]{Hunter_etal89}; 
\cite[Lisenfeld et al.\ 2002]{Lisenfeld_etal02}; 
\cite[Galliano et al.\ 2003]{Galliano_etal03}; 
\cite[B\"ottner et al.\ 2003]{Bottner_etal03};
\cite[Cannon et al.\ 2006a,b]{Cannon_etal06a};
\cite[Leroy et al.\ 2007]{Leroy_etal07}; 
\cite[Walter et al.\ 2007]{Walter_etal07}), 
and the ratio falls with metallicity 
(\cite[Lisenfeld \& Ferrara 1998]{LisenfeldFerrara98}). 
However, there is some evidence for large quantities of
cold dust that we generally miss (\cite[Galliano et al.\ 2003]{Galliano_etal03}).

One consequence of the low dust content, coupled with high porosity in some dwarfs, is
long sight-lines for stellar UV radiation. 
This heats the dust and gas in the diffuse ISM and on the surfaces
of molecular clouds. 
This results in
warmer FIR dust temperatures 
(\cite[Hunter et al.\ 1989]{Hunter_etal89}, 
\cite[Walter et al.\ 2007]{Walter_etal07}), 
and is especially an issue in starbursts 
(\cite[Madden 2000]{Madden00}). 
It also means deeper penetration
of star-forming gas clouds. This results in large photodissociation regions (PDRs) and small 
molecular cores in the star-forming clouds 
(\cite[Poglitsch et al.\ 1995]{Poglitsch_etal95},
\cite[Madden et al.\ 1997]{Madden_etal97}, 
\cite[Hunter et al.\ 2001b]{Hunter_etal01b}). 
When the molecular gas is detected, we usually find that stars are forming in
giant molecular clouds, just like in spirals
(\cite[Wilson \& Reid 1991]{WilsonReid91}, 
\cite[Ohta et al.\ 1992]{Ohta_etal92}, 
\cite[Wilson 1994]{Wilson94}, 
\cite[Taylor et al.\ 1999]{Taylor_etal99}, 
\cite[Hunter et al.\ 2001b]{Hunter_etal01b},
\cite[Leroy et al.\ 2006]{Leroy_etal06}),
but the clouds in dwarfs have a different proportion of PDR and core.
\cite[Bolatto et al.\ (1999)]{Bolatto_etal99} (Figure \ref{fig-d88})
suggest that the molecular core  in a typical cloud, as traced
by CO, shrinks and the PDR grows with decreasing metallicity.
At very low metallicities a cloud could be entirely PDR. 
And yet, the products of the star formation process---HII region luminosity functions,
cluster mass functions, and stellar initial  mass functions---appear to be normal.
Even the star formation efficiency within low metallicity clouds appear to be normal:
In NGC 6822's Hubble V star forming region the star formation efficiency is a nominal
10\% 
(\cite[Israel et al.\ 2003]{Israel_etal03}).

{\underline{\it The dust is different}}. 

In dwarfs
the mix of PAHs, large and small grains, and silicate and carbon grains 
is different than in spirals
(\cite[Madden 2000]{Madden00}, 
\cite[Hunter et al.\ 2001b]{Hunter_etal01b}, 
\cite[Engelbracht et al.\ 2005]{Engelbracht_etal05},
\cite[Hogg et al.\ 2005]{Hogg_etal05}, 
\cite[Madden et al.\ 2006]{Madden_etal06}, 
\cite[Hunter \& Kaufman 2007]{HunterKaufman07}, 
\cite[Rosenberg et al.\ 2008]{Rosenberg_etal08}).
Low metallicity ISMs are a hostile environment for PAHs:
PAHs, as well as small grains, are destroyed by intense stellar radiation fields
(\cite[Bolatto et al.\ 2000, 2007]{Bolatto_etal00}; 
\cite[Madden 2000]{Madden00}; 
\cite[Hunter et al.\ 2001b]{Hunter_etal01b}; 
\cite[Galliano et al.\ 2003]{Galliano_etal03}; 
but see 
\cite[Lisenfeld et al.\ 2002]{Lisenfeld_etal02}), 
and the presence of PAH emission
decreases as the hardness of the radiation field increases 
(\cite[Madden et al.\ 2006]{Madden_etal06}).
PAH emission relative to small grain emission correlates with metallicity
(\cite[Engelbracht et al.\ 2005]{Engelbracht_etal05}, 
\cite[Madden et al.\ 2006]{Madden_etal06}, 
\cite[Draine et al.\ 2007]{Draine_etal07};
\cite[Rosenberg et al.\ 2008]{Rosenberg_etal08};
but see \cite[Walter et al.\ 2007]{Walter_etal07}).
However, there is more PAH emission relative to starlight in higher star formation
systems 
(\cite[Hunter et al.\ 2006]{Hunter_etal06}, 
\cite[Jackson et al.\ 2006]{Jackson_etal06}).
So, there are differences in the nature of the dust in dwarfs compared to spirals, 
but does it make any difference?
There is some evidence that FIR spectral energy distributions vary among
dwarfs, but this variation may not correlate with metallicity or the intensity
of the stellar radiation field 
(\cite[Kiuchi et al.\ 2004]{Kiuchi_etal04}).

{\underline{\it The lower dust content should affect the star formation process}}. 

But does it?
Dust traces star formation, but to what extent does the lower dust content 
and different relative
dust components in dwarfs {\it affect} the star formation process?
Weak cooling of the gas should result in less efficient cloud formation because cooling takes longer
(\cite[Dib \& Burkert 2005]{DibBurkert05}).
According to 
\cite[Schaye (2004)]{Schaye04}, we can also expect a higher threshold
for star formation when the metallicity is lower.
So one would think that a lower dust and metal content would mean that a galaxy would
have a harder time forming cold clouds. This would be a particular problem in
regions of lower gas density such as the outer disk.
\cite[Calura et al.\ (2008)]{Calura_etal08} 
show how dust production and destruction in dwarfs is tied to the 
galaxy's star formation history. But, the converse must also be true.
The cooling problem would suggest that dwarfs have star formation histories in
which the star formation {\it rate} increases at first and then levels
off after the ISM reaches some metallicity.
So, we expect the lower dust content to affect the star formation process
and star formation histories,
but the connection hasn't been shown observationally.


\begin{figure}[t]
\includegraphics[width=2.5in]{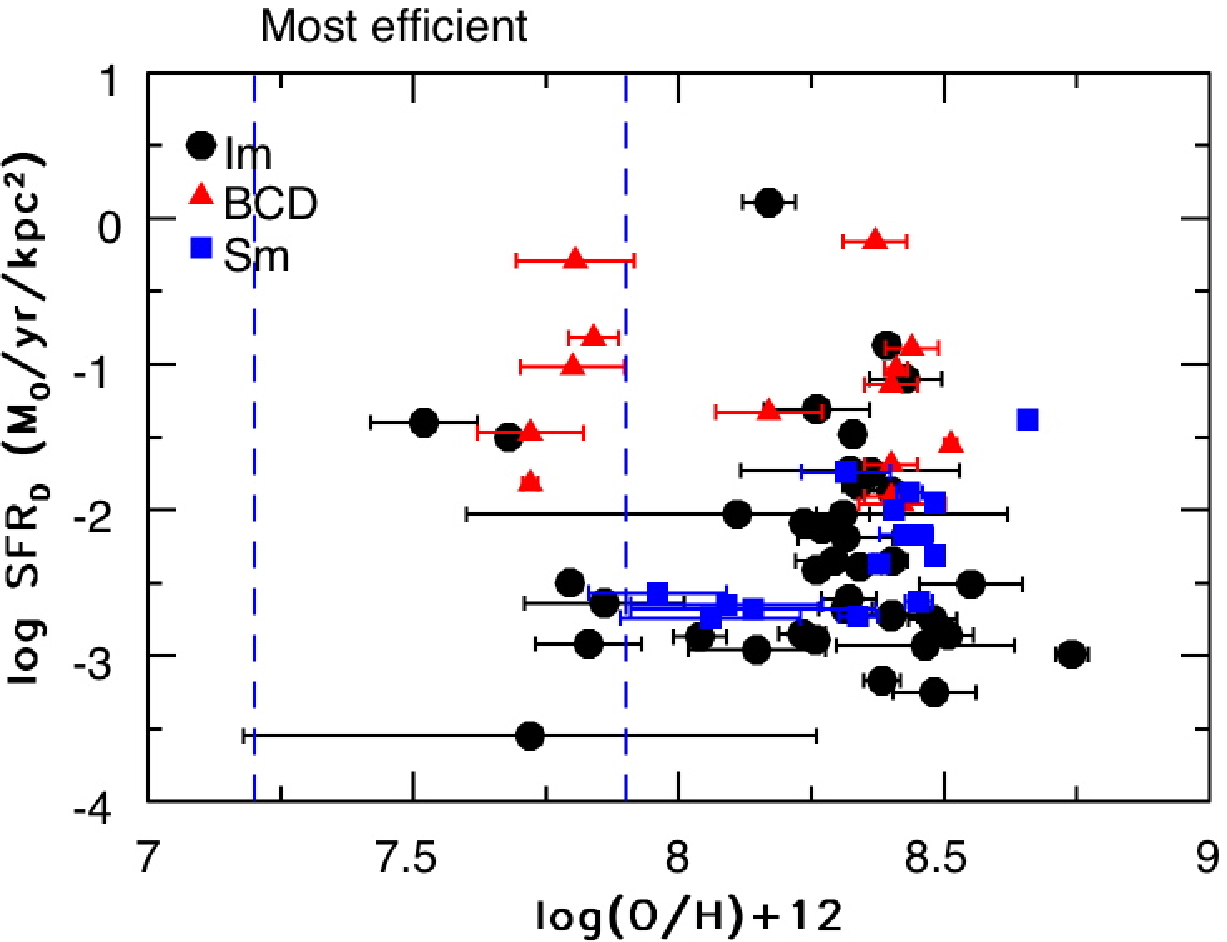}
\includegraphics[width=2.5in]{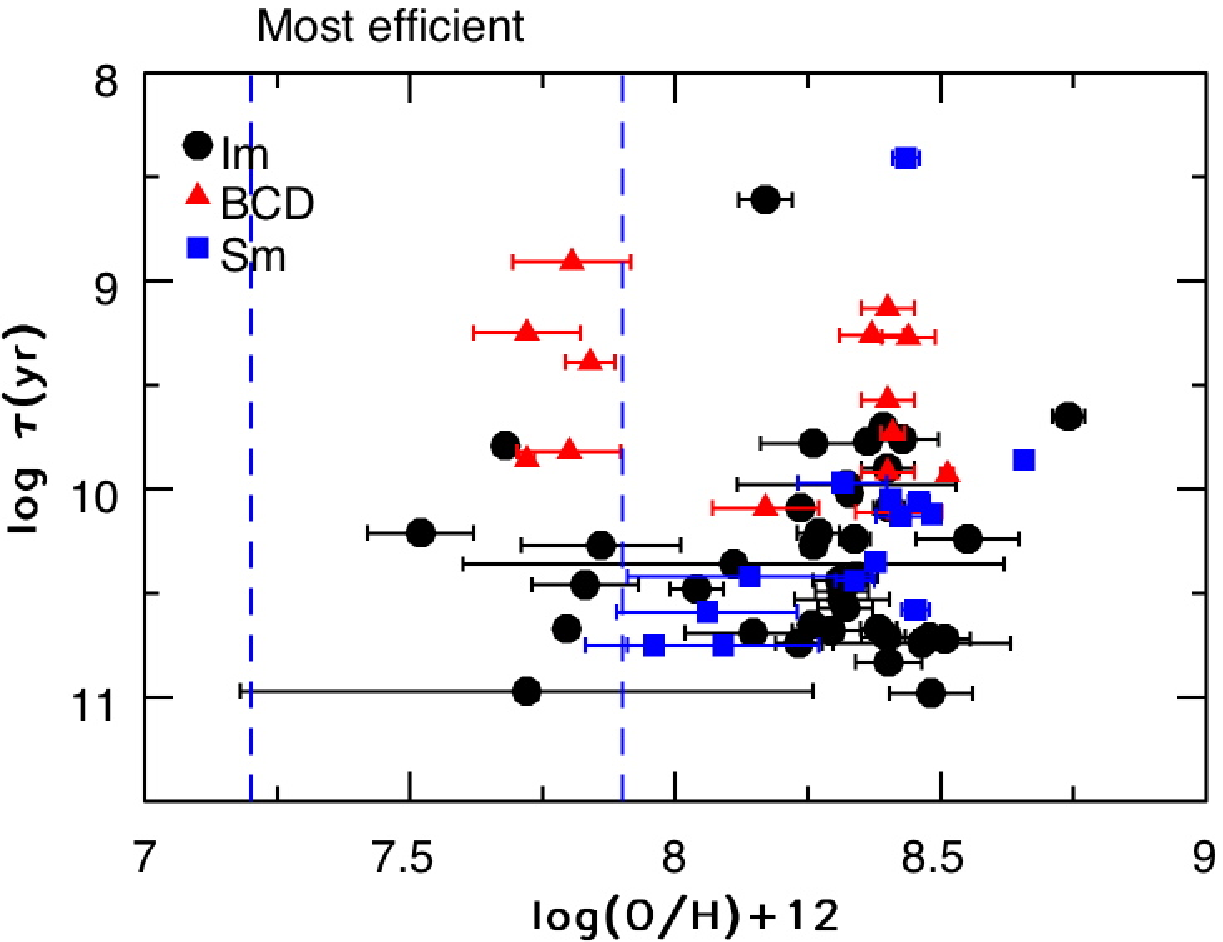}
\caption{O/H vs.\ the SFR/$\pi R_D^2$ ({\it left}) and vs.\ $\tau=M_{gas}/$SFR ({\it right}). 
The models of \cite[Spaans \& Norman (1997)]{SpaansNorman97} suggest that over the
indicated metallicity range star formation is most efficient. The galaxies with the high star
formation rates are BCDs, which may have other things going on, so 
we are not seeing an obvious effect. We need a better way to examine this prediction.}
\label{fig-ohsfr}
\end{figure}

\section{Initiating star formation}

So, what do we need to form stars: high enough gas density and cool enough gas 
temperatures. 
\cite[Spaans \& Norman (1997)]{SpaansNorman97} 
have argued that a multi-phase ISM emerges when the metallicity
reaches 0.02$Z_{\mathord\odot}$, and this is very important to the efficiency of star formation.
Locally, the lowest
metallicity star-forming galaxy has an oxygen abundance that is 1/40 $Z$$_{\mathord\odot}$
(SBS0335-052, \cite[Izotov et al.\ 2005]{Izotov_etal05}). 
Increasing metallicity enhances atomic and molecular cooling; increasing dust content
absorbs part of the stellar radiation field and boosts H$_2$, and life is good (for star
formation). Things go sour when the metallicity reaches 0.1$Z_{\mathord\odot}$ and ambipolar
diffusion for clouds is no longer important. 
Then the ease of star formation should go down. (See Figure \ref{fig-ohsfr}).

But before the first stars form and there are no metals, it would seem that cloud formation 
by normal means would be difficult 
(\cite[Li et al.\ 2005a]{Li_etal05a}). 
However, we have heard at this meeting that a single generation of very 
massive stars forming in the initial dark matter mini-haloes can raise the metallicity
of the gas to $10^{-3}$-$10^{-3.5}$$Z_{\mathord\odot}$ and that star formation enters the realm of
quasi-normal above this critical metallicity. 
So, a freshly formed dwarf galaxy would not be starting from zero metallicity.
Even so, this threshold is well below the
value of 0.02$Z_{\mathord\odot}$ where 
\cite[Spaans \& Norman (1997)]{SpaansNorman97} 
argue star formation becomes particularly easy.

I gather that some of the ``small'' galaxies in the early universe were dense objects---denser than 
dwarfs, and they became the
central regions of today's spirals (\cite[Mao, Mo, \& White 1998]{Mao_etal98}). 
Forming clouds in these units
wouldn't be particularly difficult because the gas density was so high.
But,
maybe to get things going in dwarf galaxies in the early universe (but after the first
stars have polluted the gas a little), you need an
especially high gas density (for a dwarf). Then once the metallicity is raised to
something like 0.02$Z_{\mathord\odot}$, cloud formation
becomes more efficient and can take place at lower gas densities.

\begin{figure}[t]
\begin{center}
\includegraphics[width=1.5in]{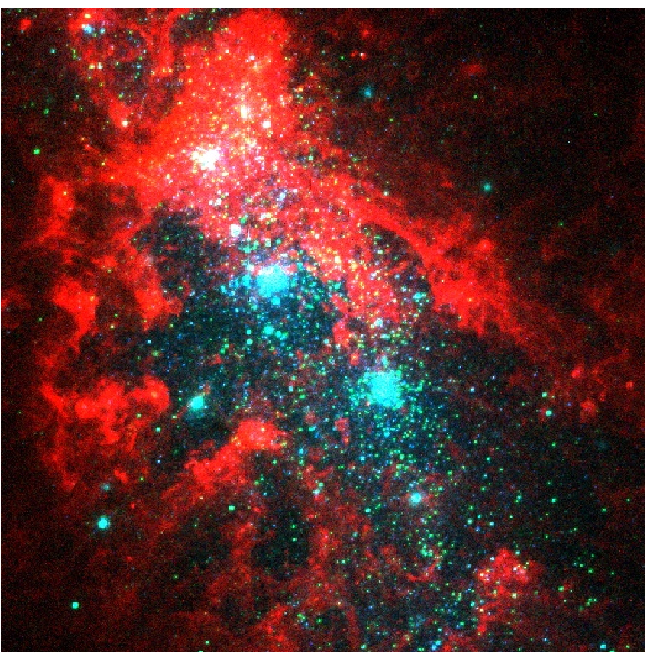}
\vspace*{0.05 cm}
\includegraphics[width=2.25in]{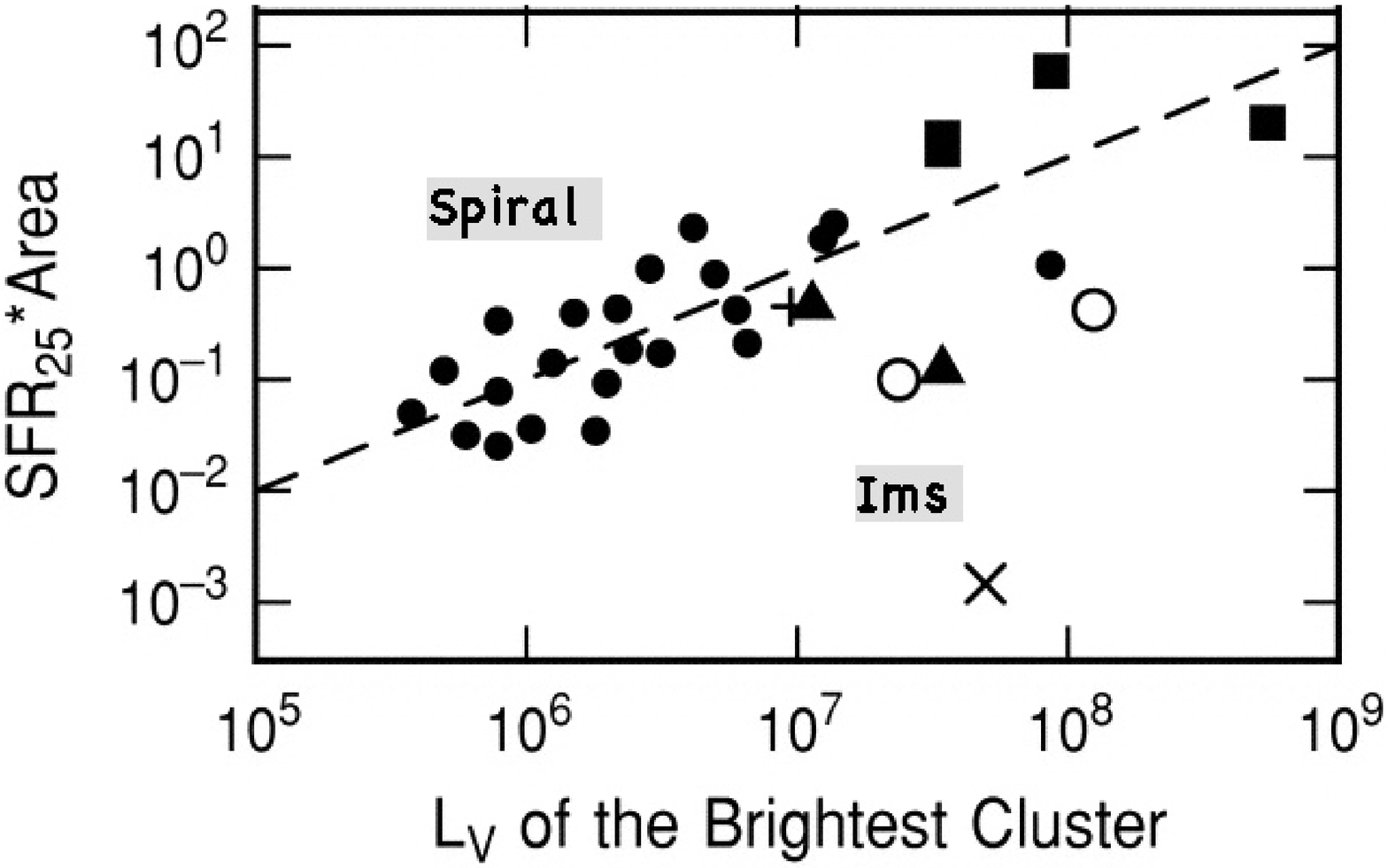}
\caption{{\it Left}: Super star clusters ({\it blue}) and ionized gas ({\it red}) in the 
starburst dwarf NGC 1569. Images are from {\it HST} observations presented by
\cite[Hunter et al.\ (2000)]{Hunter_etal00}.
{\it Right}: $V$-band luminosity of the brightest star cluster in spiral and dIm galaxies
vs. integrated galactic star formation rate. The brightest clusters in dIm galaxies are
too bright for the galaxies' star formation rates. Adapted from 
\cite[Billett et al.\ (2002)]{Billett_etal02}.
}
\label{fig-starcl}
\end{center}
\vspace*{-0.125 cm}
\end{figure}

How might we get the gas density up (temporarily) in (at least some) dwarfs?
One way is through dwarf--dwarf interactions.
Interactions, even minor ones, 
can have a big impact on a dwarf's
ability to form stars (\cite[Dong et al.\ 2003]{Dong_etal03}). 
We see that some nearby dwarfs have formed very massive star clusters that are too big 
for the number of other clusters that are present 
(see Figure \ref{fig-starcl}; \cite[Billett et al.\ 2002]{Billett_etal02}).
These systems generally also show evidence for an external perturbation.
Stellar bar structures also facilitate star formation in that barred dwarfs tend to have higher
integrated star formation rates, and these can form in an interaction
(\cite[Hunter \& Elmegreen 2006]{HunterElmegreen06}).
So, one could imagine the
first dwarfs interacting and making massive star clusters or at least getting the high 
gas density necessary to begin forming stars.

But can a dwarf--dwarf interaction/merger
make another dwarf, such as we see today? 
That this might be possible is suggested by the Local Group dIm WLM: Conditions in this galaxy were
right to form a globular cluster very early in life
(\cite[Hodge et al.\ 1999]{Hodge_etal99})
and yet today it looks like a typical dIm galaxy
(\cite[Kepley et al.\ 2007]{Kepley_etal07}).
Many BCDs may be (minor?) dwarf--dwarf mergers.
Once the current starburst is over in a BCD, would it be distinguishable from a regular dIm?
In BCDs the HI tends to be more centrally concentrated, but 
\cite[Simpson \& Gottesman (2000)]{SimpsonGottesman00}
argue that a starburst could move much of the HI back out to the edges of the optical galaxy.
This scenario would produce HI ring structures, and such are seen in some dIm today
(c.f. Figure \ref{fig-d88}, \cite[Simpson et al.\ 2005]{Simpson_etal05}).
We need simulations of dwarf--dwarf interactions in order to see what gas
densities might be achieved and what the resulting galaxy would look like for different
types of encounters.

\vspace*{-0.5 cm}
\section{Relation to star formation in early star-forming units}

So, if I wanted to look at present-day star formation processes that might be applicable
to star formation in freshly formed dwarf galaxies, where would I look? 
First, look at outer disks. 
Dwarfs are making star-forming clouds in outer disks where gas densities
are quite low on average. Star formation out there must be pretty inefficient, but it does
take place.
So, outer disks are probably the place to look to understand
the drivers for star formation in the smallest, lowest density objects in the early universe.

Second, look at dwarfs with a range of properties including metallicity, dust-to-gas ratios, 
and luminosities.  
We don't see any $10^{-3}$-$10^{-3.5}$$Z_{\mathord\odot}$ metallicity galaxies
or dust-free dwarfs today.
But, nearby dwarfs do allow us to see trends with these properties down to the most 
extreme objects like Leo T that are forming stars nearby.

Third, look at BCDs as examples of externally induced star formation proceses that may 
have been especially important in the early universe. 

\vspace*{0.5 cm}
DAH gratefully acknowledges funding from the US National Science Foundation through
grant AST-0707563 and comments by B.\ Elmegreen, J.\ Cannon, and E.\ Brinks.





\end{document}